\documentclass[pra, singlecolumn, superscriptaddress]{revtex4-2}

\usepackage{amsmath,mathtools}
\usepackage{amsfonts}
\usepackage{amssymb}
\usepackage{graphicx}
\usepackage{subfiles}
\usepackage[colorlinks,linkcolor=red,citecolor=green]{hyperref}
\usepackage{subfigure}
\usepackage[labelfont=bf,
   justification=Justified,
   format=plain]{caption}
\usepackage{graphicx}
\usepackage{dcolumn}
    \usepackage{bm}
    \usepackage{amssymb,amsmath,color}
    \usepackage{booktabs}
    \usepackage{amsthm}
    \usepackage{dsfont}
\usepackage{tcolorbox}
\usepackage{xcolor}
\definecolor{mygreen}{rgb}{0.01, 0.31, 0.59}
\definecolor{myblue}{rgb}{0.01, 0.31, 0.59}
\hypersetup{
  colorlinks=true,   
  linkcolor=myblue,  
  citecolor=mygreen, 
  urlcolor =myblue    
}

\usepackage{cases}
\usepackage{physics}
\usepackage{subfiles}
\usepackage{varwidth}
\usepackage[normalem]{ulem}
\DeclareMathOperator*{\argmax}{arg\,max}

\newcommand{\average}[1]{\left\langle #1 \right\rangle}

\begin{document}

\title{Strongly coupled fermionic probe for nonequilibrium thermometry
}
\author{Ricard Ravell Rodríguez}
\email{ricard.ra.ro@gmail.com}
\affiliation{Basque Center for Applied Mathematics (BCAM), Alameda de Mazarredo, 14, 48009 Bilbao, Spain}
\affiliation{International Centre for Theory of Quantum Technologies, University of
Gda\'{n}sk, Jana Bażyńskiego 1A, 80-309 Gda\'{n}sk, Poland}
\author{Mohammad Mehboudi}
\address{Technische Universit\"{a}t Wien, Stadionallee 2, 1020 Vienna, Austria}
\author{Micha{\l} Horodecki}
\affiliation{International Centre for Theory of Quantum Technologies, University of
Gda\'{n}sk, Jana Bażyńskiego 1A, 80-309 Gda\'{n}sk, Poland}
\author{Martí Perarnau-Llobet} 
\email{marti.perarnaullobet@unige.ch}\affiliation{Département de Physique Appliquée, Université de Genève, 1211 Genève, Switzerland}

\begin{abstract}
We characterise the  measurement sensitivity, quantified by the Quantum Fisher Information (QFI), of a single-fermionic thermometric probe strongly coupled to the sample of interest, a fermionic bath, at temperature $T$. 
For nonequilibrium protocols, in which the probe is measured before reaching equilibrium with the sample, we find new behaviour of the measurement sensitivity arising due to non-Markovian dynamics. First, we show that the QFI displays a highly non-monotonic behaviour in time, in contrast to the Markovian case where it grows monotonically until equilibrium, so that non-Markovian revivals can be exploited to reach a higher QFI. 
Second,  the QFI rate is  maximised at a finite interrogation time $t^*$, which we characterize, in contrast  to the  solution  $t^* \rightarrow 0$ known in the Markovian limit [Quantum 6, 869 (2022)]. Finally, we consider  probes make up of few  fermions and discuss different collective enhancements in the measurement precision. 
\end{abstract}
\maketitle

\section{Introduction}
Estimating the temperature of quantum systems is challenging but crucial; temperature  appears as a relevant parameter in the Gibbs state of quantum systems even if they evolve unitarily~\cite{gogolin2016equilibration,popescu2006entanglement,goldstein2006canonical}. The growing field of quantum thermometry is dedicated to developing cutting edge experimental protocols for reading out the temperature of cold systems operating at the quantum regime~\cite{mehboudi2019using,mitchison2020situ,PRXQuantum.3.040330,boeyens2023probe}. It is also devoted to investigating the fundamental bounds on achievable precision that are set by quantum physics through Cram\'er-Rao inequalities \cite{mohareview,binder2018thermodynamics}. More recently, we have witnessed an effort in making the practical and fundamental sides meet, 
with new experiments working close to the fundamental theoretical  bounds~\cite{Bouton2020,Adam2022,napolitano2011interaction,liu2021experimental,brida2010experimental,abbott2016observation} 
and a more general  characterisation of sub-optimal 
measurements in the theory side~
\cite{mohareview,binder2018thermodynamics,correa2015individual,de2016local,paris2015achieving,mehboudi2019using,Razavian2019,hovhannisyan2018measuring,mehboudi2015thermometry,hofer2017quantum,seah2019collisional,campbell2017global,mitchison2020situ,mok2021optimal,latune2020collective,rubio2021global,potts2019fundamental,jorgensen2020tight,Hovhannisyan2021,Mihailescu2023}. 

A standard  scheme involves  a small  quantum system,  used as a \textit{probe} or \textit{thermometer}, and a sample of interest at temperature $T$. After preparing the probe in a suitable state and letting it interact  with the sample, the former is measured to infer  $T$. The theory of thermometry is well understood when the probe reaches  equilibrium before being measured, a regime known as \emph{equilibrium thermometry}. Starting from the fundamental bounds on the sensing precision of~\cite{paris2015achieving}, the form of optimal probes has been characterized~\cite{correa2015individual,abiuso2022discovery}; and  the impact of  strong coupling, bath induced correlations, and sub-optimal measurements has been readily studied~\cite{correa2017enhancement,PhysRevLett.128.040502,brenes2023multispin}.  The non-equilibrium regime, i.e. when the probe is measured before reaching equilbrium, requires a more precise knowledge of the non-equilibrium dynamics but also offers new possibilities for estimating $T$ with a higher precision~\cite{Brunelli2011,correa2015individual,Landi2022,Razavian2019} and  lower backaction~\cite{Albarelli2023}. This has been extensively explored in the context of ultracold gases~\cite{Sabn2014,mehboudi2019using,mitchison2020situ,Oghittu2022,Khan2022,Yuan2023,brattegard2023thermometry}. 
The fundamental precision limits of non-equilibrium thermometry have been established in~\cite{sekatski2022optimal} when the dynamics of the probe are well described by Markovian dynamics. Beyond this regime, much less is known about the form of optimal protocols, and more generally about the impact of strong coupling and correlations  (see Ref.~\cite{brattegard2023thermometry} for recent interesting progress).  In this work, we aim to make progress in this direction by characterising  the metrological properties of a fermionic probe, make up of a single or few fermions, strongly coupled to an ultracold fermionic bath.     

We organise the paper as follows. In Section \ref{sec:setup} we introduce the resonant-level model, which  involves a single-fermion (strongly) coupled to a fermionic bath in the wide band limit. 
Next, in Section~\ref{sec:Ultimate} we discuss the error in the estimation process and corresponding  bounds. 
Section~\ref{sec:equilibrium} is dedicated to our first results regarding equilibrium thermometry with strong coupling, with emphasis in the ultracold regime. In Section~\ref{sec:nonequilibrium} we present our results for non-equilibrium thermometry, in particular by deriving corrections to the result of~\cite{sekatski2022optimal} arising due to the finiteness of the system-bath coupling. We extend our considerations to multiple-fermionic probes and discuss the impact of bath-induced interactions in Section~\ref{sec:multiple}. Finally, in Section~\ref{sec:conclude} we make some concluding remarks.

\section{Resonant level model}\label{sec:setup}
We consider a single fermionic mode strongly coupled to a fermionic reservoir, also known  as the resonant level model~\cite{schallerbook}. This model becomes exactly solvable in the wide band limit, which makes it an ideal testbed for analysing  strong coupling quantum thermodynamics~\cite{Ludovico2014,Esposito2015b,Bruch2016,Haughian2018,Pancotti2020,Tong2022,Rolandi2023,nello2023thermodynamics,Gian2023}, and here we exploit it for nonequilibrium thermometry. 

Consider a bath of non-interacting fermions that rests at thermal equilibrium with chemical potential $\mu$ at some temperature $T$. The Hamiltonian of the fermionic bath reads 
\begin{align}
    H_B = \sum_q \omega_q b^{\dagger}_q b_q,
\end{align}
where $b^{\dagger}_q$ and $b_q$ are the fermionic creation and annihilation operators for the mode $q$ in the bath with the corresponding energy $\omega_q$. 
We are interested in estimating the temperature of the Fermionic bath.
The state of the bath can thus be written as
\begin{align}
    \pi_B = \frac{e^{-(H_B-\mu N)/T}}{Z},
\end{align}
with $Z\coloneqq {\rm Tr}e^{-(H_B-\mu N)/T}$ being the partition function, and $N\coloneqq \sum_q b^{\dagger}_q b_q$.

As for the probe, we consider a single fermionic mode---extension to multiple fermionic modes is discussed later in Sec.[\ref{sec:multiple}]. Let the Hamiltonian of the probe be
\begin{align}
    H_p = \epsilon d^{\dagger} d,
\end{align}
with $d^{\dagger}$ and $d$ being the fermionic creation and annihilation operators of the probe mode with the energy $\epsilon$. The probe can be initially prepared in any suitable state. However, as we just have a single fermionic mode, due to fermionic super-selection rules the only physical states are incoherent in the number basis and should admit $\rho_p(0)=p_0(0) |0\rangle \langle 0 | + (1-p_0(0)) |1\rangle \langle 1|$ for any time, where $|0\rangle$ and $|1\rangle$ correspond to having $0$ or $1$ fermions in the mode, respectively.

In order to acquire information about the bath temperature, one allows for probe-bath interaction. To this aim we consider the following linear interaction
\begin{align}
    H_{I} = \sum_q  (t_q b^{\dagger}_q d+ t^*_q d^{\dagger} b_q).
\end{align}
Putting everything together, the total Hamiltonian reads
\begin{align}
    H = H_p + H_B + H_I,
    \label{eq:fullH}
\end{align}
and the initial state is $\rho_{PB}(0) = \rho_p(0)~\pi_B$. At any given time, the state of the probe is
\begin{align}
    \rho_p(t) = {\rm Tr}_B \left[ e^{-itH} \rho_{PB}(0) e^{-itH} \right].
\end{align} 
As already discussed, the probe's state respects the fermionic super-selection rule, and at any time it should read $\rho_p(t)=p_1(t) |1\rangle \langle 1|+(1-p_1(t)) |0\rangle \langle 0 | $. Thus, it suffices to find $p_1(t)$ to fully characterise the state at any time. This can be done by noting that 
\begin{align}
    p_1(t) = {\rm Tr}[\rho_p(t) d^{\dagger} d] = \average{d^{\dagger}d(t)},
\end{align}
where we moved into the Heisenberg picture in the last equality.

\subsection{Dynamics} 
For completeness, now we present the solution of the exact dynamics~\cite{schallerbook}, following the approach presented in~\cite{Rolandi2023,Gian2023}. 
We start from the  Heisenberg equations of motion   for $\average{d^{\dagger}d(t)}$. By using the fermionic algebra, i.e., $\{d,d^\dagger \}=\mathbb{I}$, $\{b_q,b^{\dagger}_k\} = \mathbb{I} \delta_{qk}$, and $\{b_q,b_k\}=\{d,b^{\dagger}_q\}=\{d,d\}=0$, and some straightforward calculations one finds [see \cite{schallerbook} and Appendix~\ref{app:Langevin} for details]
\begin{align}\label{eq:d_t_dot}
    \dot{d}(t) + i\epsilon d(t)  + \int_0^t \chi(s-t)d(s)ds = -i\xi(t).
\end{align}
Here, we have defined 
\begin{align}
    \xi(t) \coloneqq \sum_q t^*_q e^{-i\omega_q t} b_q(0),
\end{align}
    which can be seen as a Brownian force, and
\begin{align}
    \chi(t) \coloneqq \sum_q |t_q|^2 e^{i \omega_q t} \Theta(t)=\frac{\Theta(t)}{2\pi}\int_{-\infty}^{\infty} d\omega \Gamma(\omega) e^{i \omega t},
\end{align}
    which can be seen as a memory kernel and is responsible for memory effects. Here, the Heaviside function $\Theta(t)$ guarantees causality. Finally,
\begin{align}
     \Gamma(\omega) \coloneqq 2\pi \sum_q |t_q|^2 \delta(\omega-\omega_q),
\end{align}
     is the bath spectral density. The two time-correlation functions of the bath are often entering the equations of the motion and they read
     \begin{align}\label{eq:fdr}
         \average{\xi^{\dagger}(t) \xi(s)} & = \sum_{qq^{\prime} } t_q^* t_{q^{\prime}}e^{i(\omega_q t - \omega_{q^{\prime}}s)} \average{b_q^{\dagger}(0)b_{q^{\prime}}(0)} = \sum_q |t_q|^2 e^{i\omega_q(t-s)} f_{\beta}(\omega_q)\nonumber\\
         & = \sum_q |t_q|^2 e^{i\omega_q(t-s)} f_{\beta}(\omega_q) \int_{-\infty}^{\infty} d \omega \delta(\omega - \omega_q)  
         =\frac{1}{2\pi} \int_{-\infty}^{\infty} d \omega f_{\beta}(\omega) \Gamma(\omega) e^{i\omega(t-s)},
     \end{align}
     with the Fermi distribution defined as
     \begin{align}
         f_{\beta} (\omega)=\frac{1}{e^{\beta(\omega-\mu)}+1}.
     \end{align}
To proceed further, we restrict to the wide-band limit, i.e., we take a flat spectral density $\Gamma(\omega) = \Gamma$. It is not difficult to verify that in this limit $\chi(t) = \Gamma\delta(t)$.
The solution to Eq.~\eqref{eq:d_t_dot} reads
\begin{equation} \label{eq:d_t_single} 
d(t)=e^{(-i\epsilon-\frac{\Gamma}{2})t}d(0) - i \int_0^t ds e^{(i\epsilon+\frac{\Gamma}{2})(s-t)} \xi(s).
\end{equation}
By putting Eqs.~\eqref{eq:fdr}  and \eqref{eq:d_t_single} together, one gets
\begin{align}
p_1(t) & =\langle d^{\dagger}d (t)\rangle = e^{-\Gamma t}\average{d^{\dagger}d(0)} + \int_0^t\int_0^t~ds~ du~ e^{(i\epsilon+\frac{\Gamma}{2})(s-t)}e^{(-i\epsilon+\frac{\Gamma}{2})(u-t)}\average{\xi^{\dagger}(u)\xi(s)}
\nonumber\\
&=e^{-\Gamma t} p_1(0) + \frac{2\Gamma}{\pi}\int_{-\infty}^{\infty} d\omega f_{\beta}(\omega)\frac{1 -2 e^{-\Gamma t/2} \cos [(\omega-\epsilon)t]+e^{-\Gamma t}}{\Gamma^2+4(\omega-\epsilon)^2}, \label{eq:population-int}
\end{align}
See the appendix for details.
This integral can be evaluated numerically. Clearly, $p_1(t)$ is a function of the bath temperature, thus it can be used to infer the bath temperature. In the next section, we address how one can do so and what are the precision limits in doing that.
\subsection{The rate equation}
While our study is founded on the exact solution of the equtions of motion, here we would like present a Gorini-Kossakowski-Lindbladian-Sudarshan (GKLS) master equation \cite{gorini1976completely,lindblad1976generators} against which we contrast our main results throughout the paper. In particular, following \cite{potts2019introduction} one can find the following master equation
\begin{align}
\partial_t \rho_p(t) = -i[H_p,~\rho_p(t)] + \Gamma(1-f_{\beta}(\epsilon))D[d]\rho_p(t) + \Gamma f_{\beta}(\epsilon) D[d^{\dagger}] \rho_p(t),
\end{align}
where we defined the super operators $D[A]\circ \coloneqq A\circ A^{\dagger} - 1/2 (A^{\dagger} A \circ + \circ A^{\dagger} A)$. One can use this in the Heisenberg picture to find the so called rate equation 
\begin{align}
    \partial_t p_1(t) = -\Gamma [p_1(t) - f_{\beta}(\epsilon)],
\end{align}
that can be solved to obtain
\begin{align}\label{eq:population-rate}
    p_1(t) = e^{-\Gamma t}p_1(0) + (1-e^{-\Gamma t})f_{\beta}(\epsilon).
\end{align}

While this equation is useful in a variety of settings it should be revisited in strong coupling regime, or at small temperatures, or for short times. In our results below we see when/how the thermometry precision obtained from this approximation deviates from that of the exact solution i.e., Eq.~\eqref{eq:population-int}.

\section{Ultimate limits on estimating the temperature}\label{sec:Ultimate} 
We assume that one can always break down our thermometry protocol into three stages of (i) preparing the probe in an appropriate state $\rho(0)$, (ii) encoding the temperature into the probe via probe-bath interaction, after which the probe evolves to $\rho(t)$, and (iii) performing a measurement with POVM elements $M_x \geqslant 0$ and $\sum_x M_x = I$. Repeating this total $k$ times, one gets a dataset of outcomes ${
\bf x
}=\{x_1,\dots,x_k\}$ which one can use to assign an estimate ${\tilde T}({\bf x})$ to the temperature. The estimator is unbiased if on average is equivalent to the true temperature, that is $\sum_{\bf x} p_{\bf x}(t){\tilde T}({\bf x}) = T$, with $p_{\bf x}(t)=\Pi_{k}p_{x_k}(t)$ and $p_{x}(t) = {\rm Tr} [M_{x}\rho_p(t)]$. The Cram\'er-Rao inequality states that the error---quantified by the mean square error---of any unbiased estimator is lower bounded by the inverse of the Fisher information. That is~\cite{Cramer:107581,Rao1992}
\begin{align}\label{eq:CRB}
    \Delta^2 {\tilde T} \coloneqq {\sum_{\bf x} p_{\bf x}(t)\left({\tilde T}({\bf x}) -  T\right)^2} \geqslant \frac{1}{{\cal F}[p_{\bf x}(t)]} = \frac{1}{k{{\cal F}[p_x(t)]}},
\end{align}
with the Fisher information defined as
\begin{align}
    {{\cal F}[p_{\bf x}(t)]}\coloneqq \sum_{\bf x}  p_{\bf x}(t)\left[\partial_T \log p_{\bf x}(t)) \right]^2,
\end{align}
which is an additive quantity for repeated measurements (independent probabilities), hence the final equality on the rhs of \eqref{eq:CRB}. At the limit of large $k$, the CRB is known to be saturated, e.g., by using the maximum likelihood estimator~\cite{kay1993fundamentals,kolodynski2014precision}. As such, we take the total Fisher information ${{\cal F}[p_{\bf x}(t)]}$ as our figure of merit. 
For a two-outcome measurement, the Fisher information is simple and reads
\begin{equation}\label{eq:FI_2}
    \mathcal{F}[p_{\bf x}(t)] = \frac{\left[\partial_T p_1(t)\right]^2}{\left(1-p_1(t)\right)p_1(t   
    )}.
\end{equation}
 
Before explicitly analysing $\mathcal{F}[p_{\bf x}(t)] $, some general remarks can be make about the optimisation of $\mathcal{F}[p_{\bf x}(t)]$ for the protocols considered.  
To begin with, at stage (i), the initial state of the probe is bounded to be diagonal. Since the Fisher information is convex~\cite{holevo2011probabilistic,toth2014quantum,paris2009quantum} the optimal initial state is therefore either the ground state, or the exited state.
Furthermore, at the stage (ii) we assume the freedom in choosing how long the probe evolves before being measured. Here, we also analyse how the Fisher information depends on the coupling strength and other parameters of the system. We do so in two paradigmatic frameworks: equilibrium and non-equilibrium thermometry which we will explain in the following sections. 
Finally, at stage (iii) the measurement can be optimised to the one with the maximum Fisher information, a.k.a. the quantum Fisher information (QFI) which is officially defined as
\begin{align}\label{eq:QFI_max}
    {\cal F}^{\rm Q}[\rho_p(t)] \coloneqq \max_{\{M_x\}} {\cal F}[p_{\bm x}(t)].
\end{align}
Since the maximisation is over measurement realisations one can write down the ultimate limit, the quantum Cram\'er-Rao bound, on the error which can be potentially saturated for large enough $k$. We have:
\begin{align}
    \Delta^2 {\tilde T} \geq \frac{1}{k{{\cal F}[p_x(t)]}} \geq \frac{1}{k{{\cal F}^{\rm Q}[\rho_p(t)]}}. 
\end{align}
While the maximisation in~\eqref{eq:QFI_max} may look non-trivial, one can write down an explicit expression for the QFI in terms of the density matrix~\cite{holevo2011probabilistic,toth2014quantum,paris2009quantum}
\begin{align}\label{eq:QFI_SLD}
    {\cal F}^{\rm Q}[\rho_p(t)] & = {\rm Tr}[\rho_p(t) \Lambda^2],~~\text{where}~~\Lambda\rho_p(t)+\rho_p(t)\Lambda \equiv 2\partial_T \rho(T).
\end{align}
The observable $\Lambda$ which is defined above is actually forming the basis for the optimal measurement as well. Lastly, let us remark that in case the probe is a single fermion, the only possible measurement is the population measurement and the optimisation over all measurements is unnecessary, thus when dealing with a single fermion we simply use the Fisher information terminology, instead of the quantum Fisher information. However, in case of multiple fermions [Sec.~\ref{sec:multiple}] we use the quantum Fisher information.

\section{Equilibrium thermometry}\label{sec:equilibrium}
We first consider equilibrium thermometry, where the probe reaches a steady state due to the interaction with the sample. 
This scenario is relevant when time is not a resource; instead the total number of probing repetitions $k$ is fixed. One often assumes that the probe reaches a steady state before being measured---i.e., one measures $\rho_p(\infty)$.
In the simple case of weak interaction between the probe and the system, the probe thermalizes and its state reads $\rho_p(\infty)=\tau=e^{-\beta H_p }/\mathcal{Z}$, i.e. a Gibbs state.

In this scenario, the optimal measurements are energy measurements~\cite{correa2015individual,stace2010quantum,jahnke2011operational} and the Fisher information of the probe at equilibrium is proportional to its heat capacity~\cite{jahnke2011operational}.  On the other hand, in the case of strong interactions between the probe and the system, the steady state of the probe is described by the mean-force Gibbs state \cite{trushechkin2022open}.

\begin{figure}[h]
\centering
\includegraphics[scale=0.9]{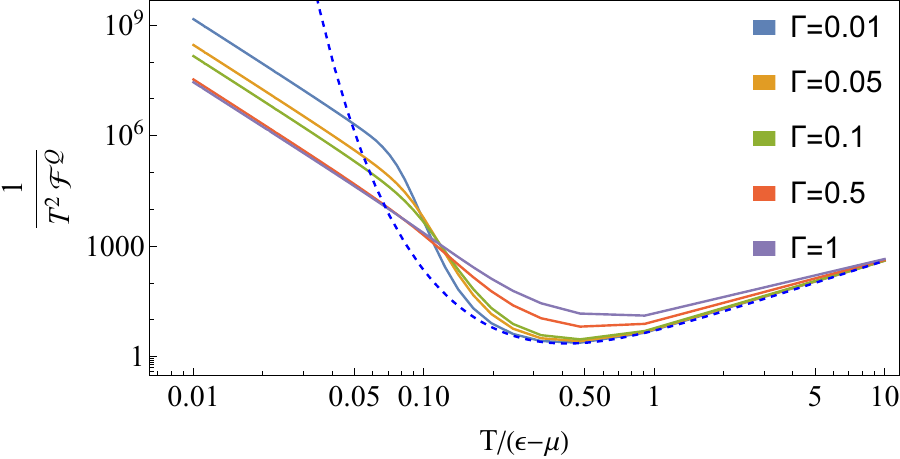}
\caption{Loglog scale plot of the the minimum relative error, $1/(T^2 {\cal F}^Q[p_x(\infty)])$ as a function of temperature $T$. Different interaction strengths $\Gamma$ are considered. For comparison, we also plot the relative error for a thermal state---that is the solution of the GKLS master equation---the dashed line.  We set $\epsilon-\mu=1$. 
}    \label{fig:error-T}
\end{figure}

For bosonic baths and probes, it has been shown that one can highly improve the precision when estimating ultralow temperatures by increasing the coupling between the system and the probe \cite{correa2017enhancement,hovhannisyan2018measuring,PhysRevLett.128.040502}. This can be understood by noting that for finite coupling, the QFI of the probe becomes sensitive to the spectrum of the full Hamiltonian (probe+sample), which is gapless. As a consequence, the relative error ($\Delta {\tilde T}/T$) when $T\rightarrow 0$ diverges polynomialy~\cite{hovhannisyan2018measuring,potts2019fundamental,jorgensen2020tight}, instead of the standard exponential growth for gapped systems~\cite{paris2015achieving}. We find a qualitatively similar behavior for fermionic systems, as explained below. 


When the probe reaches a steady state ($t\to\infty$), Eq.~\eqref{eq:population-int} reduces to 
\begin{align}
p_1(\infty) & = \frac{2\Gamma}{\pi}\int_{-\infty}^{\infty} d\omega \frac{f_{\beta}(\omega)}{\Gamma^2+4(\omega-\epsilon)^2}.
\label{generalp_1}
\end{align}
Firstly, in the weak coupling regime, one can see that 
$\lim_{\Gamma \to 0} p_1(\infty) = f_{\beta}(\omega)$ which coincides with the steady state of the GKLS rate equation~\eqref{eq:population-rate}. For this canonical state, one can find the Fisher information analytically. In equilibrium thermometry, it is common practice to study the minimum noise-to-signal ratio as a figure of merit, which by using the CRB reads
\begin{align}
    \frac{\Delta^2 T_{\min}}{T^2} \coloneqq \frac{1}{T^2 {\cal F}^Q[p_{\bf x}(\infty)] }.
\end{align}
By using Eq.~\eqref{eq:FI_2} one can find 
\begin{align}
    \frac{\Delta^2 T_{\min}}{T^2} = \frac{(1+e^{\beta(\epsilon-\mu)})^2}{(\epsilon-\mu)^2\beta^2 e^{\beta(\epsilon-\mu)}}.
\end{align}
Limiting to a scenario with a fixed effective gap $\epsilon_{\rm eff} = \epsilon-\mu>0$, one can see that the FI is optimal at some temperature $T^*\approx 0.42 \epsilon_{\rm eff}$. At higher temperatures $T\to \infty$ the noise-to-signal ratio diverges polynomially with temperature, i.e., $\frac{\Delta^2 T_{\min}}{T^2} \propto (T/\epsilon_{\rm eff})^2$. On the other hand, as $\beta\to \infty$ the error diverges exponentially, i.e., $\frac{\Delta^2 T_{\min}}{T^2} \propto (\beta \epsilon_{\rm eff})^{-2}\exp(\beta\epsilon_{\rm eff})$. This exponential divergence is  expected from canonical states with a finite [effective] energy gap. 

\begin{figure}[h]
\centering
\includegraphics[scale=0.7]{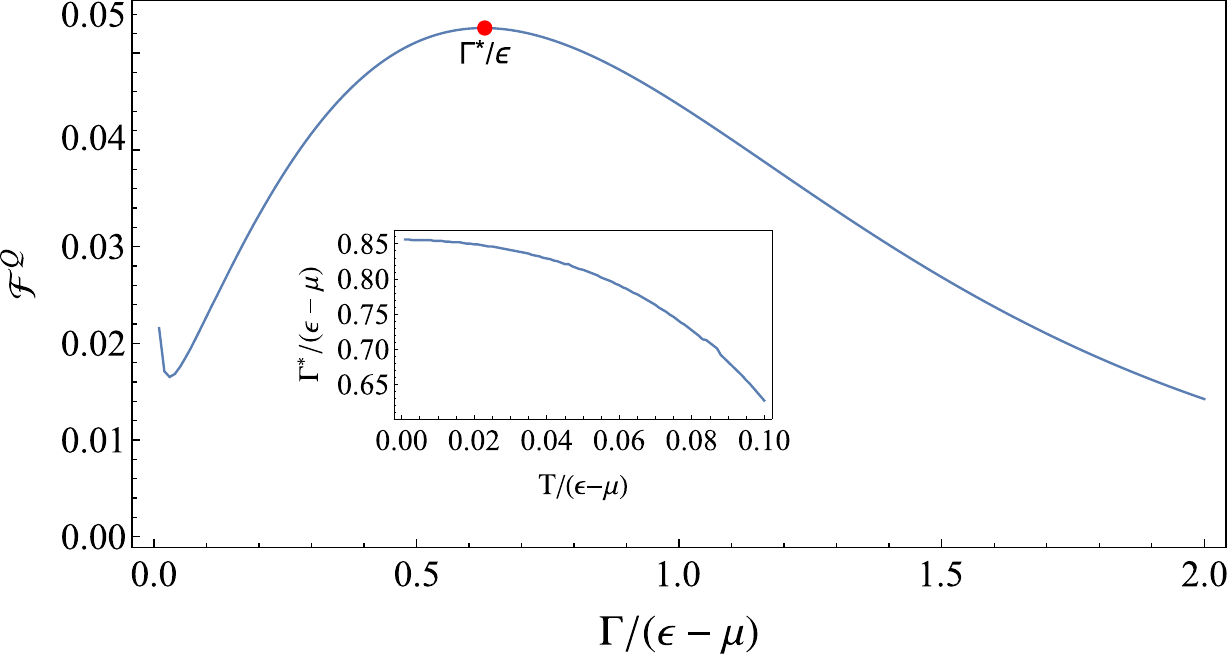}
\caption{Plot of the quantum Fisher Information  $\mathcal{F}^Q$ as a function of the coupling strength $\Gamma$ for $T/(\epsilon-\mu)=0.1$. The red point corresponds to the maximum value of the plot and it corresponds to the value obtained for the optimal coupling strength $\Gamma^*/(\epsilon-\mu)$. \textbf{Inset:} Plot of $\Gamma^*/(\epsilon-\mu)$ for the steady state as a function of the temperature $T$.  We set $\epsilon-\mu=1$ for both plots. 
}     \label{fig:fisher-vs-gamma}
\end{figure}

Moving to the strong coupling regime, we compute the QFI via the exact solution~\eqref{generalp_1}. Our results  are  illustrated in Fig.~\ref{fig:error-T}, which shows that  the error at low temperatures diverges polynomially with temperature. A numerical analysis suggests that for low enough temperatures we have $\frac{\Delta^2 T_{\min}}{T^2} \propto T^{-2-\alpha}$ where $\alpha \approx 2.05$. This behaviour is a consequence of the fact that the  spectrum of the total Hamiltonian~\eqref{eq:fullH} is gapless~\cite{hovhannisyan2018measuring,jorgensen2020tight}, and has also been observed in other models~\cite{hovhannisyan2018measuring,potts2019fundamental,jorgensen2020tight,Mihailescu2023}.
At the limit of high temperatures the behaviour coincides with the canonical distribution as the master equation becomes more reliable. Furthermore, contrary to the canonical distribution,  the exact solution is highly impacted by the coupling $\Gamma$---even though for weak coupling strengths ($\Gamma \ll \epsilon-\mu$) and for temperatures $T \approx \epsilon-\mu$ the Markovian approximation approaches the exact results, as expected. We analyse this behaviour in Fig.~\ref{fig:fisher-vs-gamma} where we fix the temperature to the low regime and depict QFI vs the coupling $\Gamma$. Evidently, the behaviour is non-monotonic, i.e., there exist an optimal $\Gamma$ which in general depends on the temperature. This is in contrast to scenarios with a bosonic probe here at low temperatures increasing the coupling monotonically improves the precision close to $T=0$~\cite{correa2017enhancement,mehboudi2019using,PhysRevLett.128.040502}. To illustrate more the dependence of the optimal coupling strength $\Gamma^*$ in the regime of low temperatures, we plot the inset of Fig. \ref{fig:fisher-vs-gamma} the optimal $\Gamma$ ($\Gamma^*$) as a function of temperature $T$. One can see that for ultralow temperatures  $\Gamma^*$ stays approximately constant. Recently, and more similarly to our result, a non-monotonic behaviour was also seen in thermometry of bosonic samples when using qubits as probes~\cite{brenes2023multispin}.

\section{Non-equilibrium thermometry}\label{sec:nonequilibrium}

We now change paradigm and consider a dynamical scenario, where one is allowed to measure the thermometer before it reaches its steady state. 

\subsection{Fisher Information}

Motivated by the fact that the Markovian approximation is not trustful for all time regimes, in this subsection we compute the Fisher Information as a function of time, for the transient dynamics of our model. Regarding the temperature regime which is more interesting to us, we see that in Fig. \ref{fig:error-T}, for temperatures small with respect to the fermionic gap, i.e. $T \ll  \epsilon-\mu$ one can see that the behaviour between the results obtained from the Markovian approximation and the analytical ones start to differ significantly. On the other hand, for $T  \gg  \epsilon-\mu$ this is no longer true and one can see that regardless of the coupling strength, the behaviours are the same in both approaches. Hence we focus on the transient behaviour when $T \ll  \epsilon-\mu$. 

\begin{figure}[h]
    \centering
    \includegraphics[scale=0.8]{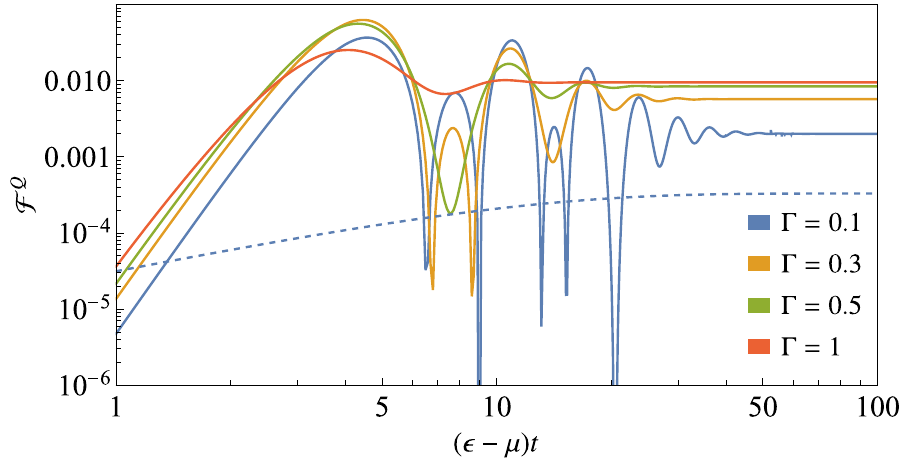}
    \caption{ Loglog plot scale of quantum Fisher Information vs time (in units of $\epsilon-\mu$) for different couplings. The dashed line corresponds to the Markovian master equation with decay rate $\Gamma=0.1$. We set $T/(\epsilon-\mu)=0.05$, $\epsilon-\mu=1$, and we start from the ground state.}
    \label{fig:fisher-vs-time}
\end{figure}

For that matter, we compute the QFI using Eq. \eqref{eq:FI_2} where the excited probability $p_1(t)$ is described in \eqref{eq:population-int}. The discrepancy can be seen in Fig. \ref{fig:fisher-vs-time} where there is a stark contrast between the Markovian and exact dynamics. In the Markovian case, the analytical formula for the quantum Fisher Information can be found by putting together Eqs.~\eqref{eq:population-rate} and \eqref{eq:FI_2} and reads
\begin{equation}    \mathcal{F}^Q=\frac{\beta^4\left(\epsilon -\epsilon  e^{-\Gamma t}\right)^2}{(2 \cosh (\beta  \epsilon )+2)^2 \left(-\frac{1}{e^{\beta  \epsilon }+1}-e^{-\Gamma  t} \left(p_1(0)-\frac{1}{e^{\beta  \epsilon }+1}\right)+1\right) \left(\frac{1}{e^{\beta  \epsilon }+1}+e^{-\Gamma t} \left(p_1(0)-\frac{1}{e^{\beta  \epsilon }+1}\right)\right)},
\end{equation}
which grows monotonically in time for any value of the parameters. Thus the Fisher Information keeps increasing until asymptotically saturating to the value corresponding to the Gibbs state for a given temperature. On the other hand, for the exact solution, one can see that there are revivals for short times which are original to non-Markovian effects. These revivals lead to a higher QFI than that obtained in the Markovian limit, whose saturation requires performing a measurement on the probe before it thermalises.
It is important to highlight that the existence of an optimal time in the exact model is  a consequence of the non-Markovian nature of the dynamics. 
This fact can be contrasted with the previous studies of Bosonic baths, where the existence of a finite optimal time is also observed~\cite{Razavian2019,cavina2018bridging}. In the bosonic case, however, the optimal time is also observed in the Lindbladian master equation approach, even for non-coherent input states~\cite{cavina2018bridging}. This is due to the fact that in the Bosonic case the rates contain some information about the temperature---which will be lost in the steady state. Instead, in our model, in the wide-band limit, the rate $\Gamma$ is independent of temperature, and thus we see the sensitivity grows monotonically in the Markovian case. Hence, the origin of the optimality of the Fisher information at finite times roots in non-Markovianity of the dynamics.

\subsection{Fisher Information rate}

In this section, we consider another figure of merit to quantify the advantages in the precision when working in the transient regime, namely the Fisher Information rate. This is useful when the equilibration time is  long or generally speaking if \textit{time} is a metrological resource. 
More precisely, we consider what is the optimal sensing strategy for a given total time $\tau$.  During this time, we 
repeat $k=\tau/t$ times ($t$ is a free parameter that can be optimised) the following procedure: probe preparation, evolution for a time $t$, and measurement. As such, the total Fisher information is equivalent to $\tau {\cal F}[p_x(t)]/t$. Putting the constant $\tau$ aside, and following Refs.~\cite{mohareview,sekatski2022optimal,mirkhalaf2022operational,correa2015individual,jevtic2015single,Pasquale2017,cavina2018bridging}, in this case our effective figure of merit will be shifted to the \textit{rate of the Fisher information} defined as 
\begin{align}\label{eq:FI_rate}
    {\tilde {\cal F}}[p_x(t)]\coloneqq \frac{{\cal F}^Q[p_x(t)]}{t}.
\end{align}
From Eq.~\eqref{eq:FI_rate} is clear that measurement time is crucial, in particular there exist an optimal measurement time. That is
\begin{align}
t^* \coloneqq \argmax_t~{\tilde{\cal F}}[p_x(t)].
\end{align}

\begin{figure}[h]
\centering
\includegraphics{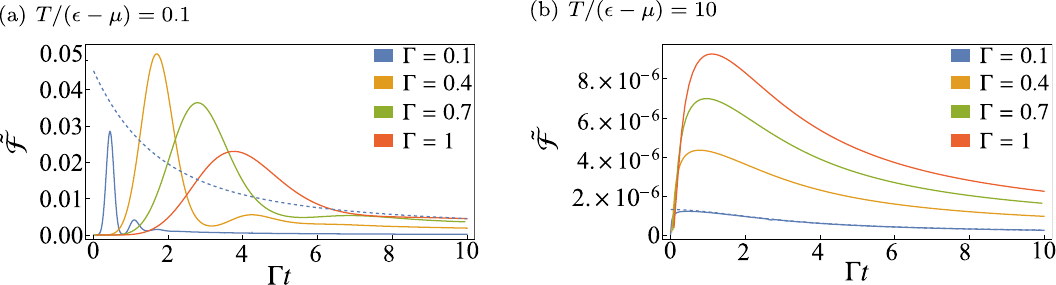}
\caption{Plot of Quantum Fisher Information over time vs $\Gamma t$ for different $\Gamma$s. The solid line corresponds to the case of Markovian Master Equation (with the decay rate being $\Gamma=0.1$).  We set $\epsilon-\mu=1$, and we start from the ground state of the dot.}
\label{fig:rate_FI_vs_t}
\end{figure}

In the weak coupling regime, we can use the GKLS master equation to calculate ${\tilde {\cal F}}[p_x(t)]$ analytically by using $p_1(t)$ from Eq.~\eqref{eq:population-rate}. For instance, if the inital state is $p_1(0)=0$ one finds
\begin{align}
    \tilde{{\cal F}}[p_x(t)] = \frac{(\epsilon-\mu)^2\beta^4(e^{\Gamma t}-1)e^{\beta(\epsilon-\mu)}}{t(e^{\beta(\epsilon-\mu)}+1)^2(e^{\beta(\mu-\epsilon)} + e^{\Gamma t})}.
\end{align}
We depict the Fisher information rate as a function of time in Fig.~\ref{fig:rate_FI_vs_t} (blue straight lines.) It can be seen that the rate is maximum at $t^*\to 0$, in agreement with previous results~\cite{correa2015individual,Pasquale2017,sekatski2022optimal}.
In fact, when using the GKLS master equation 
it has been shown that the optimal measurement time   approaches zero under quite generic weak probe-sample interactions~\cite{sekatski2022optimal}. 

Moving to the strong coupling regime,  we  exploit the exact solution~\eqref{eq:population-int} to compute numerically the rate of the FI. We  depict this in Fig.~\ref{fig:rate_FI_vs_t}. We immediately notice that the dependence of $t^*$ on the coupling and the temperature is by no means trivial. Clearly, we see from Fig.~\ref{fig:rate_FI_vs_t} that quite generically $\Gamma t^*>0$, and hence  the optimal time is strictly larger than zero, contrary to the Lindbladian master equation. 

\begin{figure}[h]
    \centering
    \includegraphics[scale=0.5]{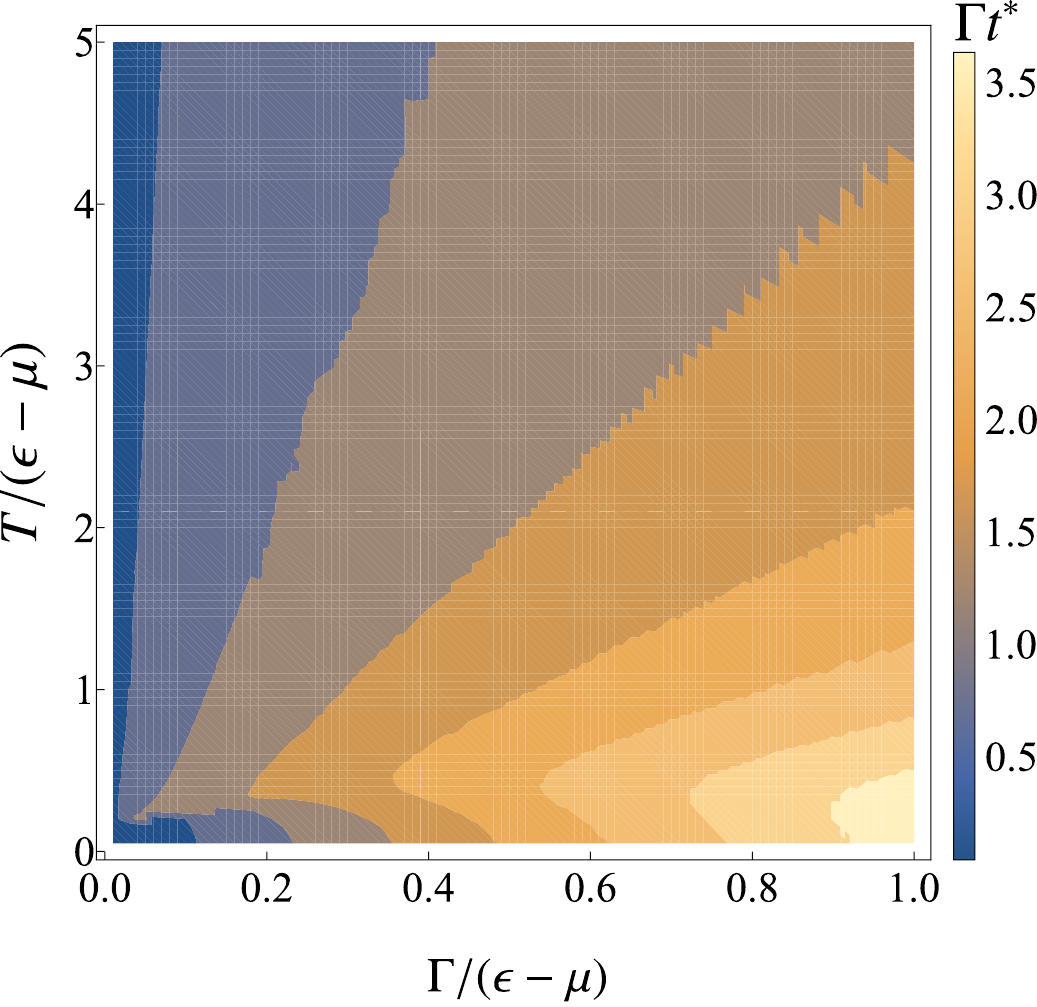}
    \caption{Contour plot of $\Gamma t^*$ as a function of $\Gamma/(\epsilon-\mu)$ and $T/(\epsilon - \mu)$. We set $\epsilon-\mu=1$, and we start from the ground state of the dot.}
    \label{fig:contour-plot}
\end{figure}

In order to elucidate more on the behavior of $t^*$ with respect to the temperature $T$ and the coupling $\Gamma$, we make a contour plot which is shown in Fig. \ref{fig:contour-plot}. We plot  $\Gamma t^*$ as a function of $\Gamma$ and $T$, where the adimensional quantity $\Gamma t^*$ is the comparison between the optimal time and the thermalization rate of the probe. As a sanity check, we see that $\lim_{\Gamma \rightarrow 0}\Gamma t^* \rightarrow 0$ which is the expected result for the weak coupling limit \cite{correa2015individual,Pasquale2017,sekatski2022optimal}. On the other hand, as the coupling $\Gamma$ increases the quantity $\Gamma t^*$  also increases which can also be seen from Fig. \ref{fig:rate_FI_vs_t}.

\begin{figure}[t]
\centering
\subfigure[$T/(\epsilon-\mu)=0.1$]{\label{max-acc-fisher-low}\includegraphics[width=80mm]{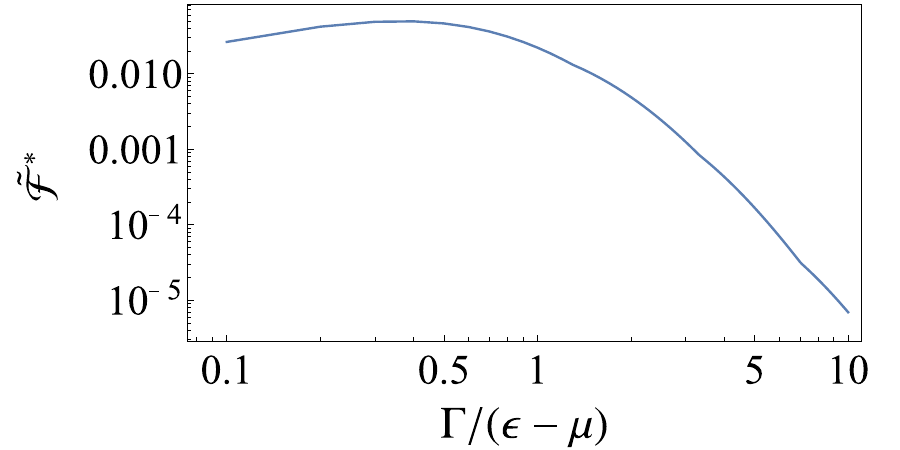}}
\subfigure[$T/(\epsilon-\mu)=10$]{\label{max-acc-fisher-big}\includegraphics[width=80mm]{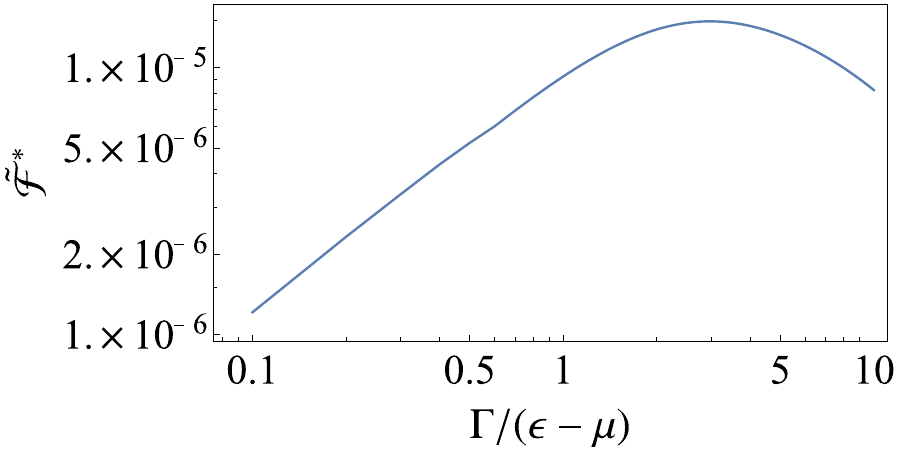}}
\caption{Loglog scale plot of the maximum value of the Fisher Information rate for different $\Gamma$s. We set $\epsilon-\mu=1$, and we start from the ground state of the dot. \label{fig:Opt_F_rate_vs_Gamma}}
\end{figure}

Finally, in Fig. \ref{fig:Opt_F_rate_vs_Gamma}, we plot the maximum value of the accumulated Fisher Information as a function of $\Gamma$ to see that the maximum value does not increase monotonically with $\Gamma$ making the behavior richer. As in the equilibrium case, there is an optimal coupling strength that maximises the QFI.

\section{Extension to multiple fermionic probes}\label{sec:multiple}

We now consider the multiple fermionic scenario where $n$ fermioins are embedded in a common bath. In this case, the total Hamiltonian reads
\begin{equation}\label{eq:n_fermions_H}
    H=\sum_{i=1}^n {\epsilon}_i d^{\dagger}_id_i + \sum_q \omega_q b^{\dagger}_q b_q + \sum_{i=1}^n \sum_q  (t_q b^{\dagger}_q d_i+ t^*_q d^{\dagger}_i b_q),
\end{equation}
where we assume the fermions are non-interacting, and their interaction with the bath is independent of their index. In the Appendix, we discuss  the  general solution to this model. Let us here  consider two insightful scenarios. 
\subsection{The symmetric case}
Firstly, we consider a case where all the fermions have the same energy, $\epsilon$. Exploiting this symmetry, we can define a new set of fermions such that only one mode is coupled to the bath. 
For instance, with two fermions the Hamiltonian can be rewritten as
\begin{equation}
    H= \epsilon d_+^{\dagger}d_+ +  \epsilon d_-^{\dagger}d_- + \sum_q \omega_q b^{\dagger}_q b_q +  \sqrt{2}\sum_q  (t_q b^{\dagger}_q d_+ + t^*_q d_+^{\dagger} b_q),
\end{equation}
where $d_{\pm}=\frac{d_1\pm d_2}{\sqrt{2}}$. On the one hand, the mode $d_-$ is not coupled to any other mode hence evolves freely. It therefore gets no information about the bath temperature. On the other hand, the coupling of the 
 $d_+$ mode to the bath is intensified by a $\sqrt{2}$ coefficient. Generalising to $N$ fermions, the model has one effective fermion evolving through the dynamics with coupling strength $\sqrt{N}\Gamma$. As such, all our previous results with a single fermion apply to this case too, simply by replacing $\Gamma\to \sqrt{N}\Gamma$. Note that in this case, the optimal measurement is performed in the basis of $d^+$, which is nonlocal in the space of the original fermions. 
Note also that 
the steady state is not unique because it depends on the initial conditions (because the expectation value for the second mode will fulfill $\langle d_-^{\dagger}d_-(t) \rangle=\langle d_-^{\dagger}d_-(0) \rangle$).

\subsection{Two fermions with different gaps}
Setting the gaps to be different, breaks down the symmetry. Here, we ask ourselves about additivity of the quantum Fisher information, i.e., how does the Fisher information of a system with two fermion in a common bath---which we denote by ${\cal F}^{\rm Q}[\rho_{12}(t)]$---compare to that of the same two fermions embedded in independent baths---which we denote ${\cal F}^{\rm Q}[\rho_{1}(t) \otimes \rho_{2}(t)]$. While the latter is additive, the former can be in general sub-additive, additive, or super-additive. Thus, we wonder 
\begin{align}\label{eq:extensivity?}
    {\cal F}^{\rm Q}[\rho_{12}(t)] \stackrel{?}{>=<} {\cal F}^{\rm Q}[\rho_{1}(t) \otimes \rho_{2}(t)].
\end{align}
Let us remark that in the common bath setting, the optimal measurement and the QFI should be chosen by solving Eq.~\eqref{eq:QFI_SLD}. This can be simply done by solving for $\rho_{12}(t)$, as explained in the Appendix.

\begin{figure}[h]
\centering
\subfigure[$T/(\Delta\epsilon-\mu)=0.1$]{\label{ratio-small-T}\includegraphics[scale=0.55]{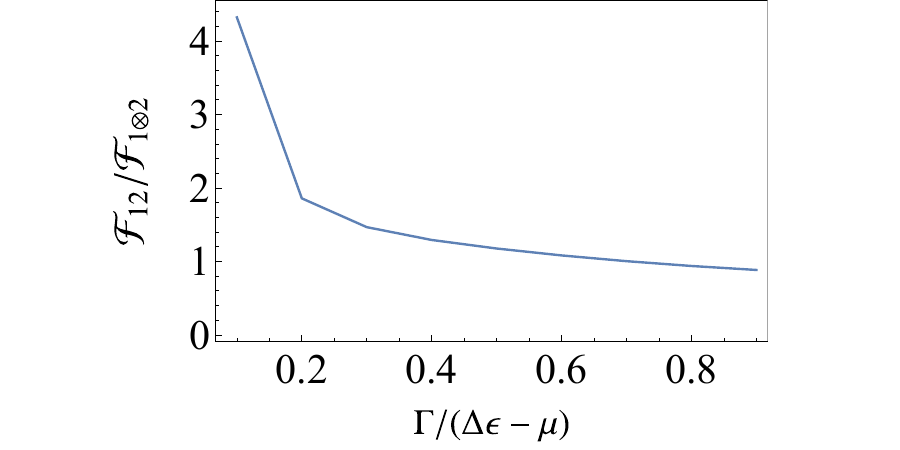}}
\subfigure[$T/(\Delta\epsilon-\mu)=100$]{\label{ratio-big-T}\includegraphics[scale=0.55]{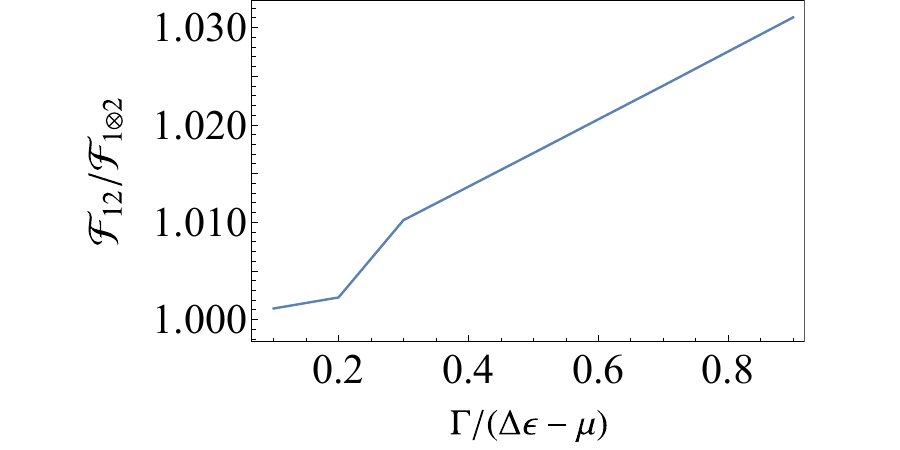}}
\caption{Loglog scale plot of  the ratio between ${\cal F}^{\rm Q}[\rho_{12}(t)] $ and $ {\cal F}^{\rm Q}[\rho_{1}(t) \otimes \rho_{2}(t)]$  ($\mathcal{F}_{12}/\mathcal{F}_{1\otimes 2}$) for different $\Gamma$'s in the steady state limit. We see that at small couplings the ratio goes to one, as expected---since the Markovian master equation holds and the steady state is thermal. However, at stronger couplings this is not the case, which is specifically more pronounced at lower temperatures. Here, we set $\Delta \epsilon - \mu =1$. \label{fig:ratio-2-fermions}}
\end{figure}

In Fig.~\ref{twofermion}, we test the additivity by depicting the two sides of Eq.~\eqref{eq:extensivity?} as a function of time. One can see that for the chosen range of parameters the QFI is initially sub-additive, but at longer times becomes super additive, i.e., ${\cal F}^{\rm Q}[\rho_{12}(t)] \geq {\cal F}^{\rm Q}[\rho_{1}(t) \otimes \rho_{2}(t)]$. To shed more light on the long time behaviour, in Fig.~\ref{fig:ratio-2-fermions} we plot the ratio between the steady state QFI in the two scenarios. Particularly, we can see that at low temperatures ($T  \ll  \Delta\epsilon-\mu$, where $\Delta \epsilon \equiv \epsilon_2 - \epsilon_1$) and for weak couplings ($\Gamma  \ll  \Delta \epsilon - \mu$) the QFI is super additive, that is the bath-induced correlations are helping to improve the precision. On the other hand, at higher temperatures, and specially at the weak coupling one sees that the two scenarios perform equal and therefore the QFI is additive. Our results agree with those obtained in \cite{brattegard2023thermometry} where the authors showed that for a system consisting of two qubits in a Fermi gas, the superadditivity of the QFI appear (so the bath correlations help in enhancing the precision) in the regimes of low temperatures---in accordance to our findings---and for low couplings between the bath and the impurities---again in agreement with our results.   

\begin{figure}[h]
    \centering
    \includegraphics[scale=0.5]{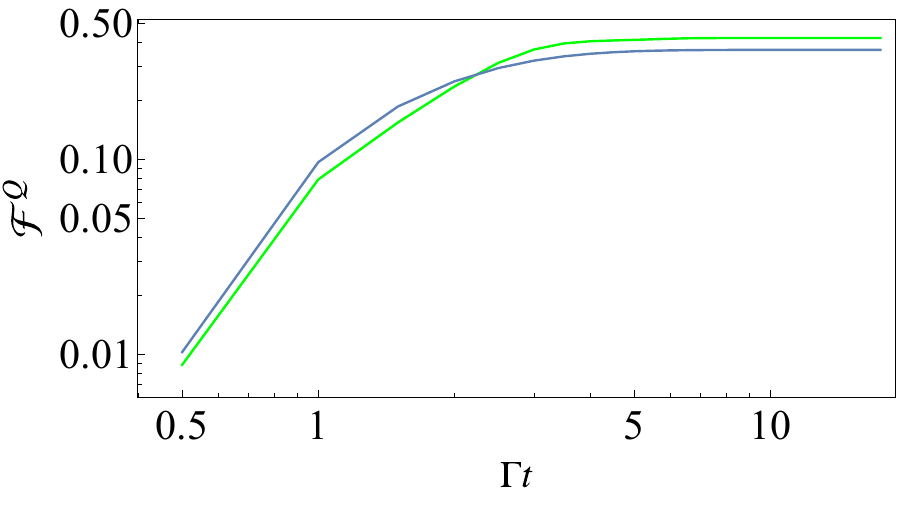}
    \caption{LogLogPlot of Quantum Fisher Information versus $\Gamma t$. Green line represents the case in which two fermions are embedded in the same bath, that is ${\cal F}^{\rm Q}[\rho_{12}(t)]$. Blue line represents the case in which each fermion is independently in contact with a fermionic bath, that is ${\cal F}^{\rm Q}[\rho_{1}(t)\otimes \rho_{2}(t)]$. Here, we set the parameters to $\Delta \epsilon - \mu =1$, $\Gamma=0.5$, $T=1$ and we start from the ground state for both fermions.}
    \label{twofermion}
\end{figure}

\newpage

\section{Conclusions}\label{sec:conclude}
Thermometry in ultracold  gases presents a formidable challenge, and recent research aims at characterising the potential of  non-invasive measurements via small  quantum probes~\cite{mohareview}. 
In this work, we characterised the thermometric potential of fermionic probes strongly coupled to a fermionic environment, by calculating their dynamics without resorting to any approximations apart from the wide-band limit. Our investigation encompasses two fundamental thermometry scenarios: equilibrium and non-equilibrium thermometry. 

In the context of equilibrium thermometry, we demonstrated that the minimum noise-to-signal ratio for low temperatures diverges polynomially when the coupling is strong, in contrast  to the standard exponential decay obtained from the weak coupling solution. A similar behaviour had been observed in other  models for strong coupling quantum thermometry, and can be understood by noting that that the full  Hamiltonian of probe+sample is gapless~\cite{hovhannisyan2018measuring,potts2019fundamental,jorgensen2020tight}. 
Our results exhibit significant sensitivity to the coupling between the bath and the probe, with the optimal interaction strength being a non-trivial function of the temperature (see Fig. \ref{fig:error-T}). 

In the transient regime, when the system has not reached equilibrium, we studied two figures of merit: the quantum Fisher Information and its rate.  For the QFI, we noted that whereas for Markovian dynamics it grows monotonically with time,  this is no longer true for the  exact non-Markovian dynamics. In the latter, we showed revivals in the QFI due to  non-Markovianity effects that can be exploited for a higher measurement sensitivity. Regarding the rate of the Fisher information, it is known that this quantity is maximised in the limit $t\rightarrow 0$ for Markovian dynamics so that the optimal protocol consists of a fast measure-and-prepare strategy~\cite{sekatski2022optimal}. Here we showed that whenever the Markovian approximation breaks down, this strategy quickly becomes suboptimal and the Fisher information rate is maximized for a finite $t^*>0$. The dependence of $t^*$ on the temperature and coupling strengh is non-trivial and shown in  Fig \ref{fig:contour-plot}. 

Additionally, we have explored scenarios involving multiple probes simultaneously immersed in the same fermionic bath, which we solved analytically. We studied superadditivity at the steady state for two specific temperatures that could be considered as low and high. At low temperatures, we observed that the Fisher Information of the probes collectively coupled to a single bath is higher than the sum of the Fisher information obtained with each fermion connected to a separate bath. 
As expected, such a phenomenon disappears at high temperatures due to the lack of correlations.

Several intriguing open questions await future exploration. For scenarios involving more than one fermion, is the experimental setup we have examined here feasible, or should we consider all fermions connected to the same bath with varying coupling strengths? Furthermore, we have exclusively considered the minimum noise-to-signal ratio, which imposes restrictions on the specific measurement that one carries. One could wonder about sub-optimal measurements in the multi-fermion case. A promising avenue for future research involves studying precision when there is uncertainty about the temperature to be estimated, using Bayesian thermometry \cite{rubio2021global} or other tighter and more realistic bounds, such as Bhattacharya or Barankin \cite{bounds}. Finally, in light of recent work \cite{boeyens2023probe}, exploring the potential of adaptive protocols to enhance the results presented here represents an interesting avenue for further investigation.

\section*{Acknowledgements}
We are thankful for fruitful discussions with Gianmichele Blasi, Patrick Potts,  Alberto Rolandi, Shishir Khandelwal and Karen Hovhannisyan. R.R.R. is also grateful to the University of Genève for the warm invitation and environment and to A. de Oliveira Junior for helping with the presentation of the figures.
M.M. acknowledges funding from the DFG/FWF Research Unit FOR 2724 ‘Thermal machines in the quantum world’. M.P.L.  acknowledges funding from the Swiss National Science Foundation (Ambizione Grant No. PZ00P2-186067). R.R.R.  acknowledges funding from the Foundation for Polish Science through IRAP project co-financed by EU within the
Smart Growth Operational Programme (contract no.2018/MAB/5). M.H. acknowledges support from  National Science Center, Poland, through grant OPUS (2021/41/B/ST2/03207), and  R.R.R. acknowledges support from  the IKUR Strategy under the collaboration agreement between Ikerbasque Foundation and BCAM on behalf of the Department of Education of the Basque Government.

\appendix
\section{Extension to multiple fermions}\label{app:Langevin}

Here, we present an extension of the resonant level model to $n$ fermions, which is based upon~\cite{Gian2023}. We consider $n$ fermions imbued in the same fermionic bath. The systems is given by
\begin{equation}\label{hamiltonian}
    H=\sum_{i=1}^n E_i d^{\dagger}_id_i + \sum_q \omega_q b^{\dagger}_q b_q + \sum_{i=1}^n \sum_q  (t_q b^{\dagger}_q d_i+ t^*_q d^{\dagger}_i b_q).
\end{equation}

Calculating the Heisenberg equations for the operators, we get the system of equations

\begin{align}\label{eq-for-d}
    i\dot{d_j}(t) &= E_j d_j(t) + \sum_q t^*_q  b_q(t), \\ \label{eq-for-b}
    i\dot{b}_q(t) &= \omega_q b_q(t) + \sum_j t_q d_j (t)
\end{align}

First, we solve the equation for $b_q$ defining a variable $v_q(t)= e^{i \omega_q  t} b_q(t)$. We get that the solution for this variable is 

\begin{equation}
    v_q(t)= -i \sum_j t_q \int_0^t ds e^{i\omega_q s}d_j(s) + v_q(0).
\end{equation}

Moving back to $b_q(t)$ and knowing that $b_q(0)=v_q(0)$ (just by the definition) we finally have

\begin{equation}
    b_q(t)= -i \sum_j t_q \int_0^t ds e^{i\omega_q (s-t)}d_j(s) + e^{-i\omega_q t} b_q(0).
\end{equation}

We now proceed injecting the solution to \eqref{eq-for-d}

\begin{align}\label{derivative}
    \dot{d_j}(t)= -iE_jd_j(t) -i\xi(t) - \sum_j \int_0^t \chi(s-t)d_j(s)ds,
\end{align}

where $\xi(t)=\sum_q t^*_q e^{-i\omega_q t} b_q(0)$ and  $\chi(t)= \sum_q |t_q|^2 e^{i \omega_q t}$. So, if we look at the last term of rhs, we can simplify it by

\begin{align}\nonumber 
     \int_0^t \sum_q |t_q|^2 e^{i \omega_q (s-t)} d_j(s)&= \int_{-\infty}^{\infty} d\omega \delta(\omega-\omega_q) \frac{2\pi}{2\pi} \int_0^t ds \sum_q |t_q|^2 e^{i \omega (s-t)} d_j(s) \\ \label{chi-eq}
     &=\frac{1}{2\pi}\int_0^t ds\int_{-\infty}^{\infty} d\omega \Gamma(\omega) e^{i \omega (s-t)}d_j(s) \underbrace{\approx}_{\Gamma(\omega)\approx \Gamma
     } \frac{1}{2\pi} \int_0^t ds \int_{-\infty}^{\infty} d\omega \Gamma e^{i \omega (s-t)}d_j(s)\nonumber\\
     &= \Gamma \int_0^t ds \delta(s-t) d_j(s)
     = \frac{\Gamma}{2}d_j(t),
\end{align}

where we have used the wide-band approximation and considered the spectral density to be flat. 

Thanks to \eqref{chi-eq}, and the properties of Delta function, namely $\int_0^t ds \delta(s-t)f(s)=1/2f(t)$, we have 

\begin{equation}
    \dot{d_j}(t)= -iE_jd_j(t) -i\xi(t) - \frac{\Gamma}{2}\sum_j d_j(t)= -i\left(E_j-i\frac{\Gamma}{2}\right)d_j(t) -i \xi(t) -\frac{\Gamma}{2}\sum_{l\neq j} d_l(t).
\end{equation}

One can write the previous equation in vectorial form as

\begin{equation}\label{vec-eq}
    \dot{\vec{d}}=A\vec{d}-i\xi(t)\mathbb{I}_v
\end{equation}

where $\mathbb{I}_v$ represents a vector make of ones, $\vec{d}=(d_1,d_2, \cdots, d_n)$ and 
\begin{equation}
A=\begin{pmatrix}
-i(E_1 - i \frac{\Gamma}{2}) & 0 & \cdots & 0\\
0& -i(E_2 - i \frac{\Gamma}{2})&\cdots & 0 \\
\vdots & \cdots & \ddots & 0 \\
0 & \cdots & \cdots & -i(E_n - i \frac{\Gamma}{2})
\end{pmatrix} 
+ \begin{pmatrix}
0 & \frac{\Gamma}{2} & \frac{\Gamma}{2} & \cdots &\frac{\Gamma}{2}\\
\frac{\Gamma}{2} & 0 & \frac{\Gamma}{2} &\cdots &\frac{\Gamma}{2} \\
\vdots & \frac{\Gamma}{2} & 0 &\frac{\Gamma}{2}  & \frac{\Gamma}{2} \\
\vdots & \vdots & \ddots & \ddots & \frac{\Gamma}{2} \\
\frac{\Gamma}{2} & \cdots & \cdots & \frac{\Gamma}{2} & 0
\end{pmatrix}.
\end{equation}

To solve \eqref{vec-eq}, one follows the same trick as before defining a variable $\vec{D}=e^{-At} \vec{d}$ to finally obtain

\begin{equation}\label{fin-analytical}
    \vec{d}=e^{At}\vec{d}(0) - i \int_0^t ds e^{-A(s-t)} \xi(s) \mathbb{I}_v.
\end{equation}

From here, we can solve \eqref{fin-analytical} by means of numerical calculations. From the components of $\vec{d}$ one can get all the information of the state. In the simplest case with one fermion ($E_1 \equiv \epsilon$), one gets
\begin{equation}  d(t)=e^{(-i\epsilon-\frac{\Gamma}{2})t}d(0) - i \int_0^t ds e^{(i\epsilon+\frac{\Gamma}{2})(s-t)} \xi(s).
\end{equation}
Assuming now that the initial state of the fermion+bath system is given by $\rho(0)=\rho_S (0) e^{-\beta(H_B-\mu N)}/Z$ and noting that $\langle b^{\dagger}_q b_{q'} (0) \rangle = \delta_{q,q'} f_{\beta}(\omega_q)$ where $f_{\beta} (\omega_q)=\frac{1}{1+e^{\beta(\omega_q-\mu)}}$ corresponds to the Fermi distribution and $\mu$ is the chemical potential of the reservoir. With this, we can calculate $\langle d^\dagger d (t) \rangle$ as
\begin{align}
p_1(t)&= \langle d^{\dagger}d (t)\rangle = e^{-\Gamma t}\langle d^{\dagger}d(0)\rangle + \int_0^t ds \int_0^t ds'e^{\frac{\Gamma}{2}(s-t+s'-t)} e^{-i\epsilon(s'-s)}\langle\xi^*(s')\xi(s)\rangle \\
& = e^{-\Gamma t}p_1(0) + \int_0^t ds \int_0^t  ds' e^{\frac{\Gamma}{2}(s-t+s'-t)} e^{-i\epsilon(s'-s)} \sum_q |t_q|^2 e^{-i \omega_q(s-s')} f_{\beta}(\omega_q)\\
&\approx e^{-\Gamma t}p_1(0) + \frac{\Gamma}{2\pi}  e^{-\Gamma t} \int_{-\infty}^{\infty} d\omega \int_0^t ds \int_0^t ds' e^{\frac{\Gamma}{2}(s+s')} e^{-i(\epsilon-\omega)(s'-s)} f_{\beta}(\omega) \\
&= e^{-\Gamma t} p_1(0) + \frac{2}{\pi}\int_{-\infty}^{\infty} d\omega \Gamma f_{\beta}(\omega)\frac{1 -2 e^{-\Gamma t/2} \cos [(\omega-\epsilon)t]+e^{-\Gamma t}}{\Gamma^2+4(\omega-\epsilon)^2}, \label{population-int}
\end{align}
where again we have used the wide-band limit. The result corresponds to Eq. \eqref{eq:population-int} from the main text.
\section{Fluctuation dissipation relation}
The real part of the memory kernel and the two time correlation functions of the bath admit a fluctuation-dissipation relation in the frequency domain. On the one hand
\begin{align}
    {\tilde C}(\omega^{\prime})\coloneqq\int_{-\infty}^{\infty}\average{\xi^{\dagger} \xi(t)}e^{i\omega^{\prime} t}dt   =2\pi f_{\beta}(-\omega^{\prime})\Gamma(-\omega^{\prime}),
\end{align}
on the other hand, we have 
\begin{align}
    {\tilde \chi}(\omega^{\prime})\coloneqq\int_{-\infty}^{\infty} \chi(t)e^{i\omega^{\prime} t}dt = \frac{1}{2}\Gamma(-\omega^{\prime}) + \frac{i}{2\pi}\int_{-\infty}^{\infty}d\omega~\frac{\Gamma({\omega})}{\omega + \omega^{\prime}},
\end{align}
and hence
\begin{align}
    {\tilde C}(\omega) = 4\pi f_{\beta}(\omega) {\rm Re~}{\tilde \chi}(\omega).
\end{align}
\section{Short times expansion}

If we expand $d(t)$ up to the second order in t, we have

\begin{equation}
    d(t) \approx d(0) + \dot{d}(0)t + \frac{1}{2}\ddot{d}(0)t^2 = d(0) +t(-iE_1d(0)-i\xi(0))+\frac{1}{2}t^2(-iE_1\dot{d}(0)-i\dot{\xi}(0)-\chi(0)d(0)),
\end{equation}

where we used the expression \eqref{derivative} and its derivative.
If we now compute the population up to the second order

\begin{equation}
    \langle d^{\dagger}d(t) \rangle= (1+iE_1t)(1-iE_1t)n_0+t^2\langle\xi^*(0)\xi(0)\rangle-E_1^2 t^2 n_0 - \chi(0)n_0 t^2 + 0(t^3)
\end{equation}

where the remaining terms are 0 because we consider that our initial state is of the form $\rho=\rho_S \tau_B$ where $\tau_B$ is the thermal state for the bath. In addition, we have

\begin{equation}
    \langle\xi^*(0)\xi(0)\rangle = \sum_q t^*_q  \sum_{q'}t_{q'} \langle b_q(0) b_{q'}(0) \rangle = \sum_q \sum_{q'} t^*_q t_{q'} f_{\beta}(\omega_q) \delta_{q,q'} = \frac{1}{2\pi} \int_{-\infty}^{\infty}d\omega \Gamma(\omega) f_{\beta}(\omega),
\end{equation}

and 

\begin{equation}
    \chi(0) = \sum_q |t_q|^2 = \int_{-\infty}^{\infty} d\omega \delta(\omega-\omega_q) \sum_q |t_q|^2 \frac{2\pi}{2\pi} = \frac{1}{2\pi} \int_{-\infty}^{\infty}d\omega \Gamma(\omega),
\end{equation}

where $f_{\beta}({\omega})$ is the Fermi distribution for temperature $1/\beta$ and frequency $\omega.$

Finally, we can write the population as

\begin{equation}\label{eq:short_time_pop}
    \langle d^{\dagger}d(t) \rangle= n_0 +  \frac{t^2}{2\pi} \int_{-\infty}^{\infty}d\omega \Gamma(\omega) (f_{\beta}(\omega)-n_0).
\end{equation}

\subsection{Short time Fisher information}
Let us assume that the initial state is a pure state with $n_0=1$. Plugging  Eq.~\eqref{eq:short_time_pop} in Eq. \eqref{eq:FI_rate} one gets
\begin{align}
    {\cal F}[p_{\bf x}(t)] \approx \frac{t^2 (\partial_T\alpha)^2}{t\alpha},
\end{align}
where we define
\begin{align}
    \alpha \coloneqq  \frac{1}{2\pi} \int_{-\infty}^{\infty}d\omega \Gamma(\omega) (1 - f_{\beta}(\omega)).
\end{align}
As a result, $\tilde{\cal F}[p_{\bf x}(t)] \propto t$ for short times. This means, it is never optimal to measure the probe immediately, there is an optimal time.


\bibliography{References}
\bibliographystyle{unsrt}
\end{document}